\documentclass[fleqn,usenatbib]{mnras}

\usepackage{newtxtext,newtxmath}

\usepackage[T1]{fontenc}
\usepackage{soul}
\usepackage{multirow}
\usepackage{tabularx}
\usepackage[singlelinecheck=false]{caption}
\DeclareRobustCommand{\VAN}[3]{#2}
\let\VANthebibliography\thebibliography
\def\thebibliography{\DeclareRobustCommand{\VAN}[3]{##3}\VANthebibliography}

\usepackage{graphicx}	
\usepackage{amsmath}	

\title[The Emergence of Nuclear Star Clusters]{EDGE: A new model for Nuclear Star Cluster formation in dwarf galaxies}

\author[E. I. Gray et al.]{Emily I. Gray,$^{1}$\thanks{E-mail: eg00451@surrey.ac.uk}
Justin I. Read,$^{1}$
Ethan Taylor,$^{1}$
Matthew D. A. Orkney,$^{2}$
Martin P. Rey, $^{3}$
Robert M. Yates,$^{4}$
\newauthor{Stacy Y. Kim,$^{5}$
Noelia E. D. No\"{e}l, $^{1}$
Oscar Agertz,$^{7}$
Eric Andersson,$^{8}$
and Andrew Pontzen$^{6}$}
\\
$^{1}$Department of Physics, University of Surrey, Guildford, GU2 7XH, UK\\
$^{2}$Instituto de Ci\`encias del Cosmos (ICCUB), Universidad de Barcelona, Mart\'{i} Franqu\`es 1, E-08028 Barcelona, Spain \\
$^{3}$Sub-Department of Astrophysics, University of Oxford, DWB, Keble Road, Oxford OX1 3RH, UK\\
$^{4}$Centre for Astrophysics Research, University of Hertfordshire, Hatfield, AL10 9AB, UK\\
$^{5}$Carnegie Observatories, 813 Santa Barbara Street, Pasadena CA 91101, USA\\
$^{6}$Department of Physics and Astronomy, University College London, Gower Street, London WC1E 6BT\\
$^{7}$Lund Observatory, Division of Astrophysics, Department of Physics, Lund University, Box 43, SE-221 00 Lund, Sweden\\
$^{8}$American Museum of Natural History, New York, NY 10024-5102, US\\
}

\date{Accepted XXX. Received YYY; in original form ZZZ}

\pubyear{2024}

\begin{document}
\label{firstpage}
\pagerange{\pageref{firstpage}--\pageref{lastpage}}
\maketitle

\begin{abstract}
Nuclear Star Clusters (NSCs) are amongst the densest stellar systems in the Universe and are found at the centres of many bright spiral and elliptical galaxies, and up to ${\sim}$40\% of dwarf galaxies. However, their formation mechanisms, and possible links to globular clusters (GCs), remain debated. This paper uses the EDGE simulations -- a collection of zoom-in, cosmological simulations of isolated dwarf galaxies -- to present a new formation mechanism for NSCs. We find that, at a gas spatial and mass resolution of ${\sim}3\,$pc and ${\sim}161$\,M$_\odot$, respectively, NSCs naturally emerge in a subset of our EDGE dwarfs with redshift-zero halo masses of $\rm{M}_{\rm{r}200\rm{c}} \sim 5 \times 10^9$\,M$_\odot$. These dwarfs are quenched by reionisation, but retain a significant reservoir of gas that is unable to cool and form stars. Sometime after reionisation, the dwarfs then undergo a major (${\sim}$1:1) merger that excites rapid gas cooling, leading to a significant starburst. An NSC forms in this starburst that then quenches star formation thereafter. The result is a nucleated dwarf that has two stellar populations with distinct age: one pre-reionisation and one post-reionisation. Our mechanism is unique for two key reasons. Firstly, the low mass of the host dwarf means that NSCs, formed in this way, can accrete onto galaxies of almost all masses, potentially seeding the formation of NSCs everywhere. Secondly, our model predicts that NSCs should have at least two stellar populations with a large ($\gtrsim$1 billion year) age separation. This yields a predicted colour magnitude diagram for our nucleated dwarfs that has two distinct main sequence turnoffs. Several GCs orbiting the Milky Way, including Omega Centauri and M54, show exactly this behaviour, suggesting that they may, in fact, be accreted NSCs.
\end{abstract}




\section{Introduction}
Nuclear Star Clusters (NSCs) are dense stellar systems with stellar masses and half-light radii in the range $\rm{M}_\star \sim 10^{5-8}\,{\rm{M}}_\odot$ and $\rm{r}_{1/2} \sim 1-30$\,pc, respectively (see \citealt{neumayer2020nuclear} for a review). They are most common within the centres of intermediate-mass galaxies (${\sim}90$\% of $\rm{M}_{\star} \sim 10^{9.5} \rm{M}_{\odot}$ galaxies) but are also found in up to $\sim$ 40\% of dwarf galaxies (DGs; defined here as galaxies with stellar mass $\rm{M}_\star <10^9$\,M$_\odot$ \citealt{collins2022observational}), with an occupation fraction that decreases with the mass of the host galaxy \citep{sanchez2019next, carlsten2021nuc, hoyer2021nucleation}. The common co-existence of these clusters with supermassive black holes (SMBHs) suggest that they could be promising sites for the formation of SMBH seeds \citep[e.g.][]{ferrarese2006fundamental}. And, through the mergers of stellar remnants, they  are also crucibles for the formation of gravitational waves \citep[e.g.][]{antonini19}.

Despite their ubiquity, the formation channels of NSCs remain uncertain. One key proposal is that NSCs in low-mass galaxies form from the in-fall and mergers of globular clusters (GCs; \citealt{tremaine75}). This is motivated by the observed similarities between massive GCs and NSCs (specifically, their stellar mass and luminosity \citealt{cote2006acs, mclaughlin1999efficiency, den2014hst}). More massive GCs also have more rapid dynamical friction timescales, and this injection of less-enriched stars into galactic centres may provide an explanation for the origin of metal-poor nuclei that are common in low-mass galaxies (e.g. \citealt{spengler2017virgo}). This formation scenario is further supported by observations of early-type galaxies that show a lack of GCs within their inner regions, implying that these have fallen into the centre \citep{lotz2001dynamical, capuzzo2009globular}. Additionally, there is a proven correlation between the frequency of nucleated early-type galaxies and GCs \citep{miller2007globular, lim2018globular, sanchez2019next}. 

However, GC mergers alone are not sufficient to explain the late-time star formation observed in the NSCs of more massive galaxies ($\rm{M}_\star \gtrsim 10^{9} \rm{M}_{\odot}$; \citealt{walcher2006stellar, seth2006clues, kacharov2018stellar, seth2008coincidence,fahrion2022nuclear}), 
including the Milky Way \citep{paumard2006two, lu2008disk, feldmeier2015kmos,  nguyen2019improved} and M31 \citep{bender2005hst, georgiev2014nuclear,carson2015structure}. For these NSCs at least, some mechanism for further gas accretion and star formation is required. Proposals include gas accretion and cooling in major mergers \citep{mihos1994dense,milosavljevic2004origin,hopkins2010massive,guillard2016new,brown2018nuclear}, gas accretion along a non-axisymmetric stellar bar \citep{shlosman1990fuelling}, tidal forces causing gas compression \citep{emsellem2008formation} and/or clumpy star formation at high redshifts \citep{bekki2006dissipative, bekki2007formation}. The need for late time star formation in these NSCs is further supported by $N$-body simulations of GC mergers \citep{capuzzo2008merging}. These are able to explain some features of NSCs, such as their mass and size, but cannot account for their observed rotation that also implies at least some dissipation and in-situ star formation \citep{hartmann2011constraining,tsatsi2017rotation,seth2006clues, seth2008rotating, carson2015structure, nguyen2019improved}. 

It is likely that some mix of gas accretion and GC mergers (or similar) is required to reproduce the full population statistics of NSCs, as is supported by semi-analytic models \citep[e.g.][]{gnedin2014co,antonini15}. The mass of the host galaxy likely dictates which of these mechanisms are dominant in each case, such that lower mass galaxies favour GC in-fall, where galaxies > $10^9$M$_{\odot}$ are better explained by the in-situ star formation scenario \citep{neumayer2020nuclear, fahrion2021diversity, fahrion2022nuclear, fahrion2022disentangling}. However, the question of where the GCs come from in the first place is uncertain. Additionally, despite some high-resolution cosmological simulations reporting NSCs in dwarf galaxies (eg. \citealt{gutcke2022lyra, brown2018nuclear}), the emergence of GC-like and/or NSC-like objects from a fully cosmological simulations remains understudied.

In this paper, we use simulations of a $\rm{M}_{\rm{r}200\rm{c}}$ $\sim 5 \times 10^9\,{\rm{M}}_\odot$\footnote{M$_{\rm{r}200\rm{c}}$ is the mass within the virial radius (r$_{200\rm{c}}$) which is defined as the radius at which the mean enclosed density of the dwarf equals 200 $\times$ the critical density of the Universe.} dwarf galaxy, drawn from the `Engineering Dwarfs at Galaxy formation's Edge' (EDGE\footnote{\url{https://edge-simulation.github.io/}}) project, to present a new mechanism for the formation of NSCs. These dwarfs are more massive than those previously studied in EDGE, reaching a post-reionisation mass that has previously been found to stimulate renewed star formation \citep{rey2020edge}. We find that, at this mass scale, a new phenomenon can occur: self-quenching by stellar feedback following a major-merger driven starburst. The formation of a NSC is a natural result of this starburst. To fully understand this new NSC formation mechanism, we consider three ``genetically modified" versions of the same galaxy, engineered to have the major merger earlier and later. We also resimulate the same galaxy with a different random number seed to study the impact of stochastisity (that impacts both when stars form, and when they explode) on the starburst and the final galaxy properties. We show that in all cases, except where the host galaxy is not fully quenched by reionisation, an NSC forms with two distinct stellar populations with distinct ages. We explore the observational properties of our simulated NSCs, including their luminosity, size, chemistry, and kinematics, and we discuss how these predictions compare to the latest data for nearby NSCs and GCs.

This paper is organised as follows. In \S\ref{sec:method}, we describe the EDGE suite and the specific simulations that we study in this work. In section \S\ref{sec:results}, we present the results from our simulations including the emergence of NSCs in EDGE, how they form, and their observational properties. In section \S\ref{sec:discussion}, we discuss the broader implications of these findings for both GC and NSC formation. Finally, in \S\ref{sec:conclusions} we present our conclusions.

\section{Method}\label{sec:method}

We study an isolated dwarf galaxy of virial mass M$_{\rm{r}200\rm{c}}$ = 5.79×10$^{9}$ M$_{\odot}$ at redshift $z=0$, drawn from a suite of cosmological high-resolution zoom simulations: EDGE (introduced in \citealt{agertz2020edge}). All of the EDGE simulations presented here are run for a full Hubble time assuming a $\Lambda$CDM cosmology with cosmological parameters as determined by the Planck satellite mission: $\Omega_{m}=0.309$, $\Omega_{\Lambda}=0.691$, $\Omega_{\rm{b}}=0.045$, and $\rm{H}_{0}=67.77 \rm{kms}^{-1}\rm{Mpc}^{-1}$  \citep{collaboration2020planck}.

EDGE utilises the adaptive mesh refinement hydrodynamics code, \texttt{RAMSES} \citep{teyssier2002cosmological}, to track the evolution of stars, dark matter, and gas. We have a maximum spatial resolution of 3\,pc and a mass resolution of
$\rm{M}_{\rm{gas}}$ = 161$\rm{M}_{\odot}$, $\rm{M}_{\rm{\star}}$ = 300$\rm{M}_{\odot}$ and $\rm{M}_{\rm{DM}}$ = 945$\rm{M}_{\odot}$ for the fiducial simulations. The galaxy formation model implements star formation, stellar feedback, cosmic reionisation and gas cooling and heating. Cosmic reionisation is modelled as a time-dependent source of heating spread uniformly across our simulation box  \citep{haardt1995radiative, rey2020edge}.

Star formation is governed by a Schmidt Law \citep{schmidt1959rate}, shown in Equation \ref{schm}:

\begin{equation} \label{schm}
    \dot{\rho_{\star}} = \varepsilon_{\rm{ff}} \frac{\rho_{\rm{g}}}{\rm{t}_{\rm{ff}}} \text{ for } \rho_{\rm{g}} > \rho_{\rm{SF}}  \text{ and } \rm{T}_{\rm{g}} < \rm{T}_{\rm{SF}}
 \end{equation}

\noindent
where $\dot{\rho_{\star}}$ is the star formation rate (SFR),  $\varepsilon_{\rm{ff}}$ is the star formation efficiency per free-fall time, $\rho_{g}$ is the gas density in the cell, $\rm{t}_{\rm{ff}}$ is the local gas free-fall time, $\rho_{\rm{SF}}$ is the density threshold for star formation, $\rm{T}_{\rm{g}}$ is the gas temperature in the cell, and $\rm{T}_{\rm{SF}}$ is the maximum allowed temperature for star formation. $\rho_{\rm{SF}}$ and $\rm{T}_{\rm{SF}}$ are set to $300\rm{m}_{\rm{H}}\rm{cm}^{-3}$ (the density at which the molecular fraction reaches 50\%) and 100\,K, respectively. $\rm{t}_{\rm{ff}}$ is calculated as $\sqrt{3\pi / 32 \rm{G} \rho}$ and $\varepsilon_{\rm{ff}}$ is fixed to 10\%, as per \citet{grisdale2019observed}. Schmidt's law is applied stochastically via random Poisson sampling to each cell that satisfies the star formation criteria (e.g. \citealt{rasera2006history, dubois2008onset}). This determines the number of stars a cell can form in each simulation timestep, thus ensuring the amount of star particles are kept to a manageable level (see \citealt{agertz2013toward} for a detailed explanation of the procedure we adopt). Each star particle is initialised with a mass of $\sim 300$\,M${_{\odot}}$ that represents a single stellar population.

Due to the high spatial and mass resolution of the EDGE simulations, we can track the impact of individual supernovae on their surrounding interstellar medium accurately \citep{kimm2015towards}, without the need for delayed cooling (e.g. \citealt{stinson2006star}), additional momentum injection, or similar \citep{read2016dark, agertz2020edge}. At each timestep, if a star's mass is above 8 M$_{\odot}$, it has a probability of undergoing core-collapse due to the random sampling of Equation 6 in \citealt{agertz2013toward}. This ensures a continuous but discrete injection of energy, momentum, and metals from supernova, as described in \citet{agertz2013toward} and \citet{agertz2020edge}. Energy from supernovae is injected purely thermally, if the cooling radius is well resolved ($>90\%$ of the time). If not, some additional momentum is also injected to compensate for overcooling (see \citealt{agertz2020edge}). Alongside supernovae, we account for stellar winds from massive stars (5M$_{\odot}$) through continuous injection. We do not explicitly model radiative transfer (photo-ionization or photo-heating) from young stars. For a more in-depth description of the physics and sub-grid model employed in the EDGE simulations, see \citet{agertz2020edge,rey19,rey2020edge,orkney2021edge}.

\begin{figure*}
\includegraphics[width=\textwidth]{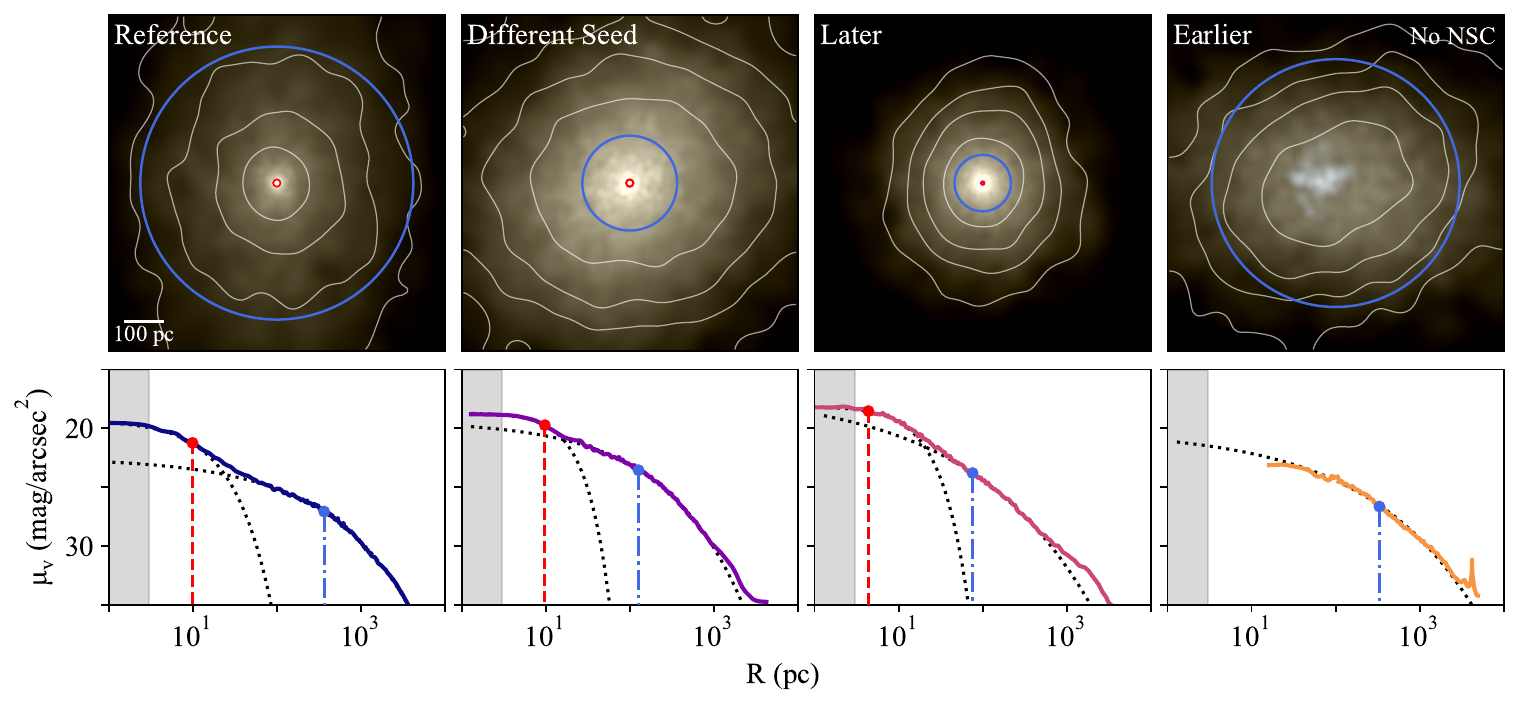}
  \caption{The emergence of NSCs in EDGE. {\bf Top:} Surface brightness contours ($\mu_v$) overlay RGB images of the stars in our simulated dwarfs at redshift $z=0$. A dense NSC is visible at the centres of the Reference, Different Seed, and Later simulations, but not the Earlier one. The images are aligned such that the angular momentum vector of the gas within 5\,kpc corresponds to the $z$-axis that points towards the reader. Surface brightness contours are in the range 18-38 mag/arcsec$^{2}$. The RGB images span values in the range 23-28 mag/arcsec$^{2}$ in which the red, blue and green channels are weighted by the I, V and U-bands, respectively. The blue and red circles denote the total stellar population half-light radius and NSC half-light radius, respectively. {\bf Bottom:} Surface brightness profiles of all the dwarfs at redshift $z=0$. The black dotted curves show double S\'{e}rsic profile fits used to obtain the half-light radii. These are marked by the red dashed and blue dash-dotted vertical lines, respectively (see also Table \ref{tab_dwarf_nsc}). The grey band to the left of each plot marks the simulation spatial resolution of 3pc. Notice that the NSC is visible in the surface brightness profiles of Reference and Different seed as a high central surface brightness, with a point of inflection. For the Later simulation, the high central surface brightness still indicates the presence of an NSC, but the inflection is now less pronounced. The Earlier simulation has no NSC and its central surface brightness profile is lower than in the other simulations.}
  \label{fig:KDE_SB}
\end{figure*}
\begin{table*}
\caption{\label{tab_dwarf_nsc} Key information about the EDGE-simulated dwarfs and their NSCs, including the half-light radius from the S\'{e}rsic fits (see Figure \ref{fig:KDE_SB}), the virial radius (of the dwarf), masses, and the average [Fe/H] and [O/Fe]. For the dwarfs, the masses and metallicities are within the virial radius whereas for the  NSCs this is within the 4 $\times$ half-light radius. The v-band magnitude within 4 times the half-light radius is added for the NSCs. The redshift the major merger occurs and its total mass ratio in comparison to the main halo is recorded.}

\begin{tabularx}{0.92\textwidth}{cccccccccc}
\hline
Dwarf & 

\begin{tabular}[c]{@{}c@{}}r$_{1/2}$\\ (pc)\end{tabular} & 
\begin{tabular}[c]{@{}c@{}}r$_{200}$\\ (pc)\end{tabular} & 
\begin{tabular}[c]{@{}c@{}}M$_{\rm{tot}}$ (<$\rm{r}_{200}$)\\ (M$_{\odot}$)\end{tabular}  & 
\begin{tabular}[c]{@{}c@{}}M$_{\star}$ (<$\rm{r}_{200}$)\\ (M$_{\odot}$)\end{tabular} & 
\begin{tabular}[c]{@{}c@{}}M$_{\rm{gas}}$ ($\rm{<}r_{200}$)\\ (M$_{\odot}$)\end{tabular}  & 
\begin{tabular}[c]{@{}c@{}}{[}Fe/H{]}\\ (dex)\end{tabular} & 
\begin{tabular}[c]{@{}c@{}}{[}O/Fe{]}\\ (dex)\end{tabular} & 
\begin{tabular}[c]{@{}c@{}}Merger \\ redshift\end{tabular} & 
\begin{tabular}[c]{@{}c@{}}Merger \\ ratio\end{tabular} \\ \hline

Reference & 365.4 $\pm$ 4.4 & 3.8 $\times$ 10$^{4}$ & 5.8 $\times$ 10$^{9}$ & 3.2 $\times$ 10$^{6}$ & 3.0 $\times$ 10$^{7}$ & -2.0 $\pm$ 0.4   & 0.6 $\pm$ 0.2 & 2.33  & 0.93   \\

Diff. Seed & 127.0 $\pm$ 2.4 & 3.7 $\times$ 10$^{4}$ & 5.6 $\times$ 10$^{9}$ & 9.6 $\times$ 10$^{6}$ & 5.1 $\times$ 10$^{5}$  & -1.8 $\pm$ 0.3  & 0.5 $\pm$ 0.2  & 2.33  & 0.91 \\

Later & 75.8 $\pm$ 1.3 & 3.7 $\times$ 10$^{4}$  & 5.3 $\times$ 10$^{9}$ & 3.3 $\times$ 10$^{6}$  & 2.0 $\times$ 10$^{7}$  & -2.0 $\pm$ 0.5  & 0.7 $\pm$ 0.1 & 1.27  & 0.36 \\

Earlier & 331.8 $\pm$ 8.8 & 3.8 $\times$ 10$^{4}$  & 5.7 $\times$ 10$^{9}$  & 4.4 $\times$ 10$^{6}$  & 5.5 $\times$ 10$^{7}$  & -2.0 $\pm$ 0.6   & 0.5 $\pm$ 0.3   & 3.35  & 0.62 \\ \hline

NSC   & 

\multicolumn{2}{c}{\begin{tabular}[c]{@{}c@{}}r$_{1/2}$\\ (pc)\end{tabular}} & 
\begin{tabular}[c]{@{}c@{}}M$_{\rm{tot}}$ (<4$\rm{r}_{1/2}$)\\ (M$_{\odot}$)\end{tabular} & 
\begin{tabular}[c]{@{}c@{}}M$_{\star}$ (<4$\rm{r}_{1/2}$)\\ (M$_{\odot}$)\end{tabular} & 
\begin{tabular}[c]{@{}c@{}}M$_{\rm{gas}}$ (<4$\rm{r}_{1/2}$)\\ ($\rm{M}_{\odot}$)\end{tabular} & 
\begin{tabular}[c]{@{}c@{}}{[}Fe/H{]}\\ (dex)\end{tabular} & 
\begin{tabular}[c]{@{}c@{}}{[}O/Fe{]}\\ (dex)\end{tabular} & 
\multicolumn{2}{c}{\begin{tabular}[c]{@{}c@{}}M$_{\rm{v}}$ (\textless{}4r$_{1/2}$)\\ (mag)\end{tabular}}  \\ \hline

Reference & \multicolumn{2}{c}{9.9 $\pm$ 1.3}  & 3.1 $\times$ 10$^{5}$  & 2.8 $\times$ 10$^{5}$  & 5.2 & -1.9 $\pm$ 0.1 & 0.5 $\pm$ 0.1  & \multicolumn{2}{c}{-8.06}  \\

Diff. Seed & \multicolumn{2}{c}{9.8 $\pm$ 0.2} & 1.4 $\times$ 10$^{6}$  & 1.3  $\times$ 10$^{6}$ & 50   & -1.9 $\pm$ 0.1 & 0.4 $\pm$ 0.1 & \multicolumn{2}{c}{-9.64} \\

Later & \multicolumn{2}{c}{4.4 $\pm$ 0.8}  & 7.7 $\times$ 10$^{5}$ & 6.9 $\times$ 10$^{5}$  & 53   & -2.0 $\pm$ 0.5  & 0.6 $\pm$ 0.1  & \multicolumn{2}{c}{-9.23} \\ \hline

\end{tabularx}
\end{table*}

\begin{figure*}
  \begin{center}
  
  \includegraphics[width=\textwidth]{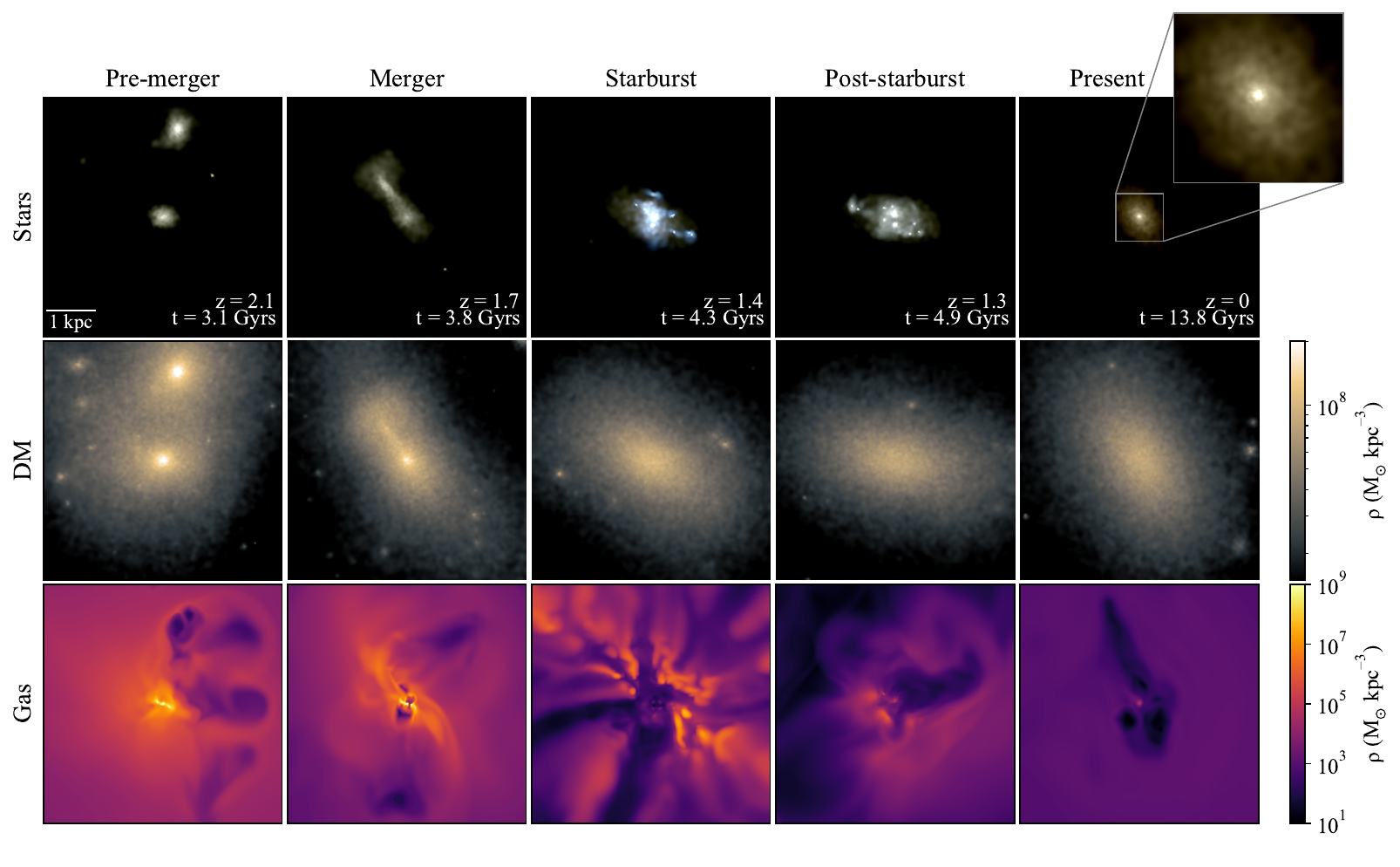}
  \end{center}
  \caption{A new mechanism for forming NSCs in low mass dwarf galaxies. The panels show the stars (top row), dark matter mass density (middle row), and gas mass density in a 0.1\,kpc slice along the $z$-direction (bottom row) of the Reference simulation for multiple snapshots centred on the main halo. The stars are shown using RGB images in the range 23-28 mag/arcsec$^{2}$ in which the red, blue and green channels are weighted by the I, V and U-bands, respectively. The panels show the major merger event which is representative of all the simulations that form an NSC. As the $\sim$ 1:1 merger merges with the main halo, the gas trapped from reionisation quenching is pushed above the star formation threshold, triggering a giant starburst -- visible as bright blue stars. At the same time, this drives a significant gas blowout (see the middle column). This starburst forms small clusters that fall to the galaxy's centre, forming the NSC.}
  \label{fig:mugshots}
  \end{figure*}

The \texttt{GenetIC} software \citep{stopyra2021genetic} provides the initial conditions for these galaxies, while also implementing a genetic modification framework that allows us to investigate different mass accretion histories for each galaxy within cosmic variance \citep{roth2016genetically, rey2018quadratic}. We change the density peak of the main halo progenitor at redshift $z=4$, but ensure the mean density of the associated Lagrangian region is kept constant. This allows us to modify the halo mass at around reionisation while keeping halo mass at redshift $z=0$ fixed. 

Haloes are selected using the \texttt{HOP} halo finder \citep{eisenstein1998hop} from a `void' simulation (a 512$^3$ resolution dark matter only simulation which covers a 50 $\rm{Mpc}^3$ patch of the universe) to be simulated at a higher resolution \citep{katz1993hierarchical, onorbe2014zoom,agertz2020edge}. To ensure selected haloes are isolated, the virial radius ($\rm{r}_{200}$) is used to find pairwise distances. Halo data are read and analysed using a combination of the \texttt{PYNBODY} \citep{pontzen2013pynbody} and \texttt{TANGOS} \citep{pontzen2018tangos} packages. 

The four simulated dwarfs utilised in the paper are described in Table \ref{tab_dwarf_nsc}. Our initial unmodified simulated dwarf is referred to as `Reference'. We engineered this dwarf to begin forming later and earlier using the genetic modification framework. `Later' refers to the dwarf that started its assembly history at a later epoch (overdensity of -20\% at z=4, and +20\% at z=0), whereas `Earlier' refers to a dwarf that had longer to form before reionisation (overdensity of +30\% at z=4, and no modification at z=0). Additionally, to study the effects of stochasticity, we rerun the unmodified simulation with the same initial conditions but a different random number seed (that impacts both when stars form and when they explode; see \S\ref{sec:method}). This dwarf is referred to as `Different Seed'.

\section{Results}\label{sec:results}

\subsection{The Emergence of NSCs in EDGE}\label{sec:nuc-emerge}

Figure \ref{fig:KDE_SB} shows the surface brightness contours of the simulated dwarfs, providing an insight into the structure of the stellar population of the dwarfs and their NSCs at $z=0$. The Reference, Different Seed and Later simulations contains a dense nucleus (or NSC) - this is not seen for the Earlier simulation. The structural differences between the Reference and Different Seed simulations show how stochasticity impacts the size and stellar mass of our simulated NSCs. The NSC size is in good agreement between these runs but the stellar mass is different by a factor of five, while the host galaxy size also differs by a factor of three.

The cylindrically averaged radial surface brightness profiles of each simulation are shown in the bottom panels of Figure \ref{fig:KDE_SB} - the alignment of the dwarf does not affect the surface brightness profile. Two S\'{e}rsic profiles were fit to the nucleated dwarfs to provide an estimate for both the NSC and host galaxy half-light radii, with uncertainties. These values are given in Table \ref{tab_dwarf_nsc}. This double-S\'{e}rsic profile is typical of real dwarf galaxies containing NSCs \citep{cote2006acs}. A single S\'{e}rsic profile was fit to the non-nucleated Earlier simulation. The Later simulation has the smallest NSC radius and reaches the highest surface brightness. In this simulation, it is less clear at what radius we transition from the NSC to the host dwarf galaxy. This difficulty in separating the NSC from its host is also seen in some real nucleated dwarfs \citep{cote2007acs}.

\subsection{The formation of NSCs via a self-quenching starburst}\label{sec:nucform}

\begin{figure}
	\includegraphics[width=\columnwidth]{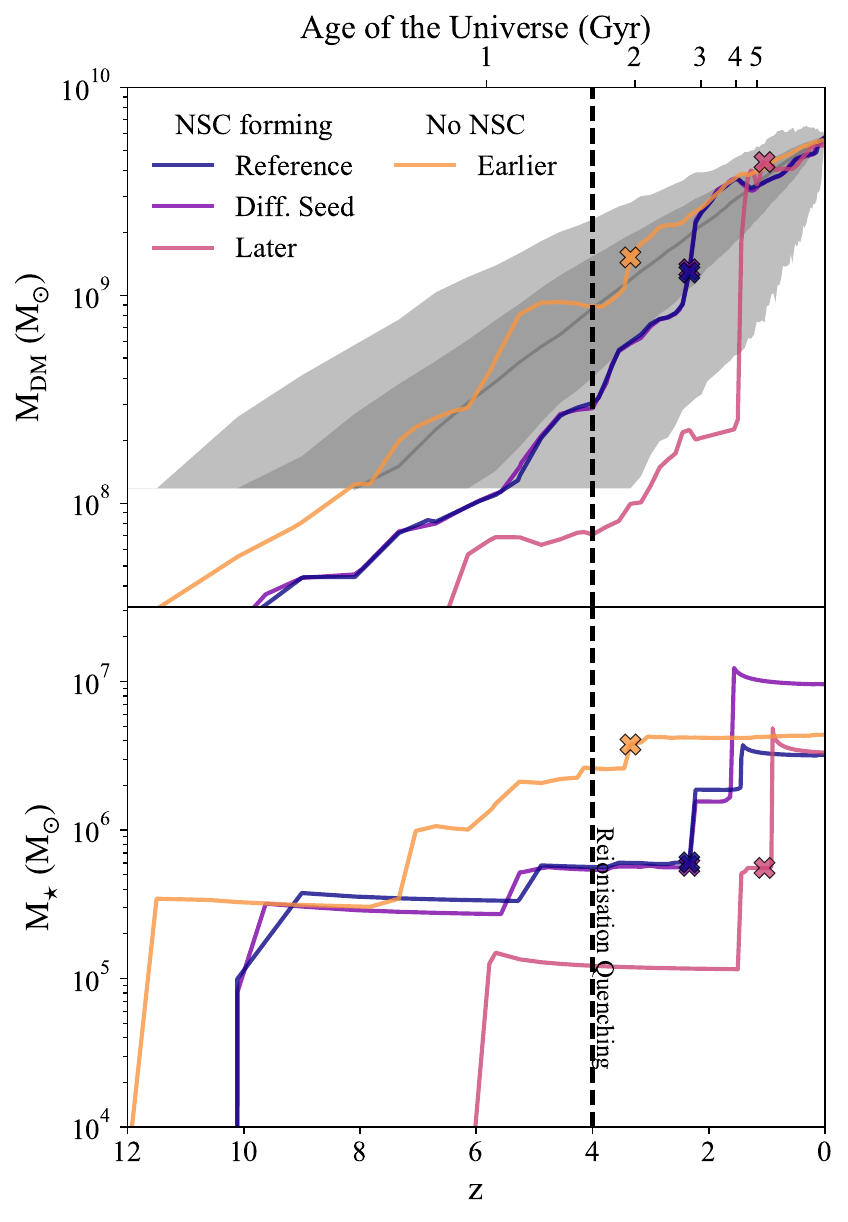}
    \caption{{\bf Top panel:} The evolution of the total DM mass within the virial radius for all of the simulations (see the legend and Table \ref{tab_dwarf_nsc} for a description of each simulation). The grey bands in the top panel show the 68\% and 95\% confidence intervals of the mass assembly trajectories of all subhaloes in the full EDGE volume. {\bf Bottom panel:} The evolution of the total stellar mass within the virial radius for all the simulations. The crosses indicate the time the r200c of the major merger crossed that of the host galaxy. The vertical dashed line marks the epoch of reionisation quenching.}
    \label{fig:dm_mass}
\end{figure}

\begin{figure}
	\includegraphics[width=\columnwidth]{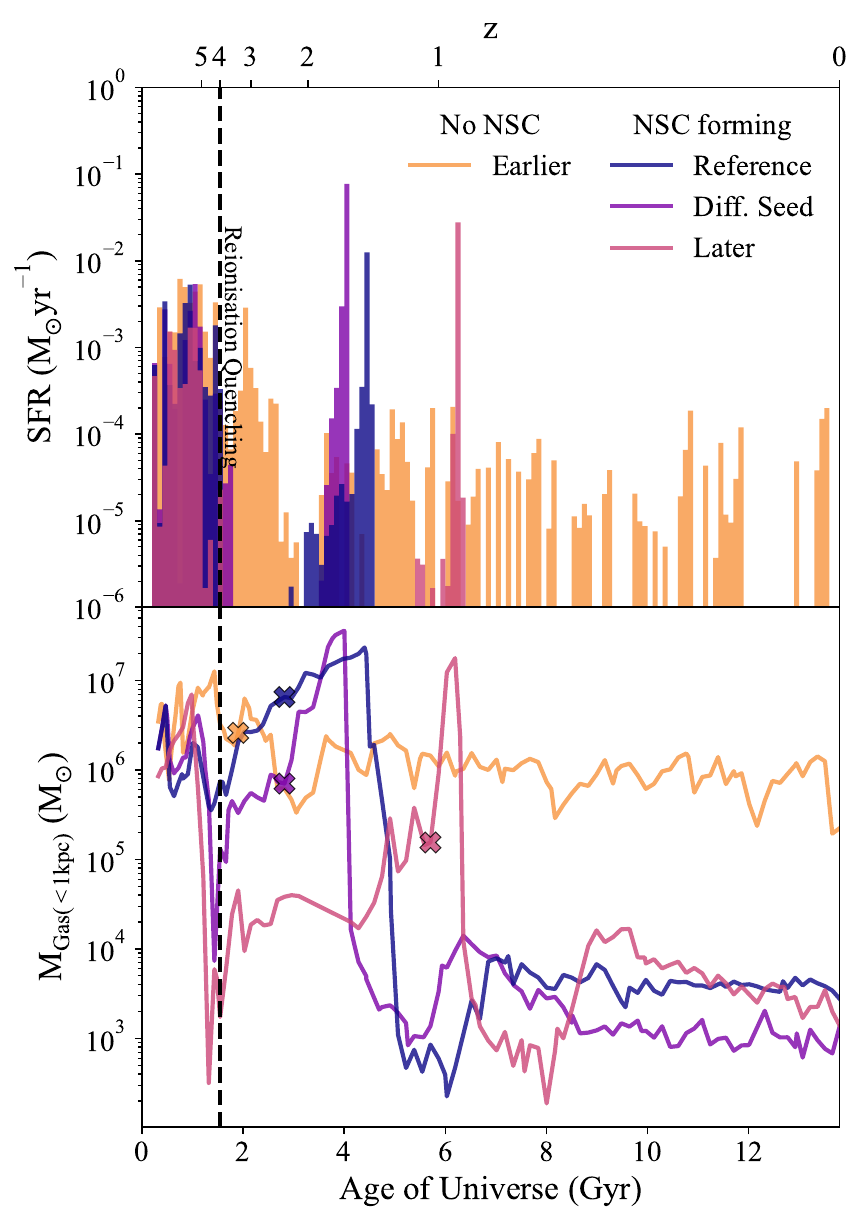}
    \caption{{\bf Top panel:} Star Formation History. {\bf Bottom panel:} Gas mass evolution within 1\,kpc of the dwarf centre. Crosses indicate when the major merger event occurs. The Reference, Different Seed and Later simulations all have early star formation which then quenches due to reionisation, leaving a trapped reservoir of gas that does not form stars. A significant merger event then triggers a starburst which self-quenches the galaxy. By contrast, the Early simulation is massive enough before reionisation that it is able to form stars continuously.}
    \label{fig:sfh}
\end{figure}

In Figure \ref{fig:mugshots}, we show graphically how an NSC forms in our Reference simulation. From top to bottom, the rows show images of stars, dark matter density and gas density (in a 0.1\,kpc thick slice along the $z$-direction) across multiple snapshots around the major merger event which triggers the NSC-forming starburst. In the leftmost panels, the two proto-galaxies (the main halo and the major merger, a $\sim$1:1 merger for this simulation, see Table \ref{tab_dwarf_nsc}) that will eventually form the final dwarf are shown. These experienced early star formation that was quenched by cosmic reionistion and have since built up their gas content, but not yet to a point where their star formation can reignite (see Rey et al. 2020, Figures 3 and 4). Prior to the starburst, the main halo has M$_{\rm{gas}}$ $\sim$ 8 $\times$ 10$^{6}$ M$_{\odot}$, while the major merger has $\sim$ 3 $\times$ 10$^{5}$ M$_{\odot}$. As can be seen in the second column, $\sim$ 2 billion years after reionisation quenching, the main halo begins to merge with the companion of similar total mass. This compresses the substantial gas reservoir in both, exciting rapid cooling and a significant starburst. The NSC forms in this starburst from a mix of smooth and clumpy star formation, producing dense star clusters (middle panels) that eventually merge into the galaxy's centre, joining the NSC (upper right panel). This starburst event is so sudden and strong that it exhausts all the remaining star-forming gas, self-quenching the galaxy. 

\subsubsection{The impact of assembly history and stochasticity}\label{sec:assembly}

To help us understand our new NSC formation mechanism further, we study the effects of different assembly histories, and of stochasticity, using our full simulation suite. Figure \ref{fig:dm_mass} shows the build up of DM (top) and stellar (bottom) mass in each of our simulations. Firstly, notice that the dark matter assembly histories for the Reference and Different Seed simulations almost perfectly overlap, as expected. But, they have different stellar mass growth curves (compare the blue and purple lines in the top and bottom panels). This difference in stellar mass occurs due to a different random number seed both when forming stars and when the most massive stars explode (see section \S\ref{sec:method}). The results are particularly sensitive for the second burst, driving a difference in the final stellar mass of a factor of $\sim 3$ just due to these stochastic effects (see Table \ref{tab_dwarf_nsc}). Nonetheless, the formation of a NSC occurs in both cases. 

Furthermore, the Reference and Earlier assembly histories lie within the 68\% confidence intervals of expected assembly trajectories: they are common (compare the blue and orange lines in Figure \ref{fig:dm_mass} to the dark grey band). By contrast, the Later simulation lies outside the 95\% intervals and is, therefore, rarer. All of the simulations were designed to reach the same final dark matter mass by $z=0$ but have very different final stellar masses due to their distinct assembly histories, similar to what has been reported for lower mass EDGE dwarfs in \citet{rey19}.

\begin{figure*}
    \centering
    \includegraphics[width=0.8\textwidth]{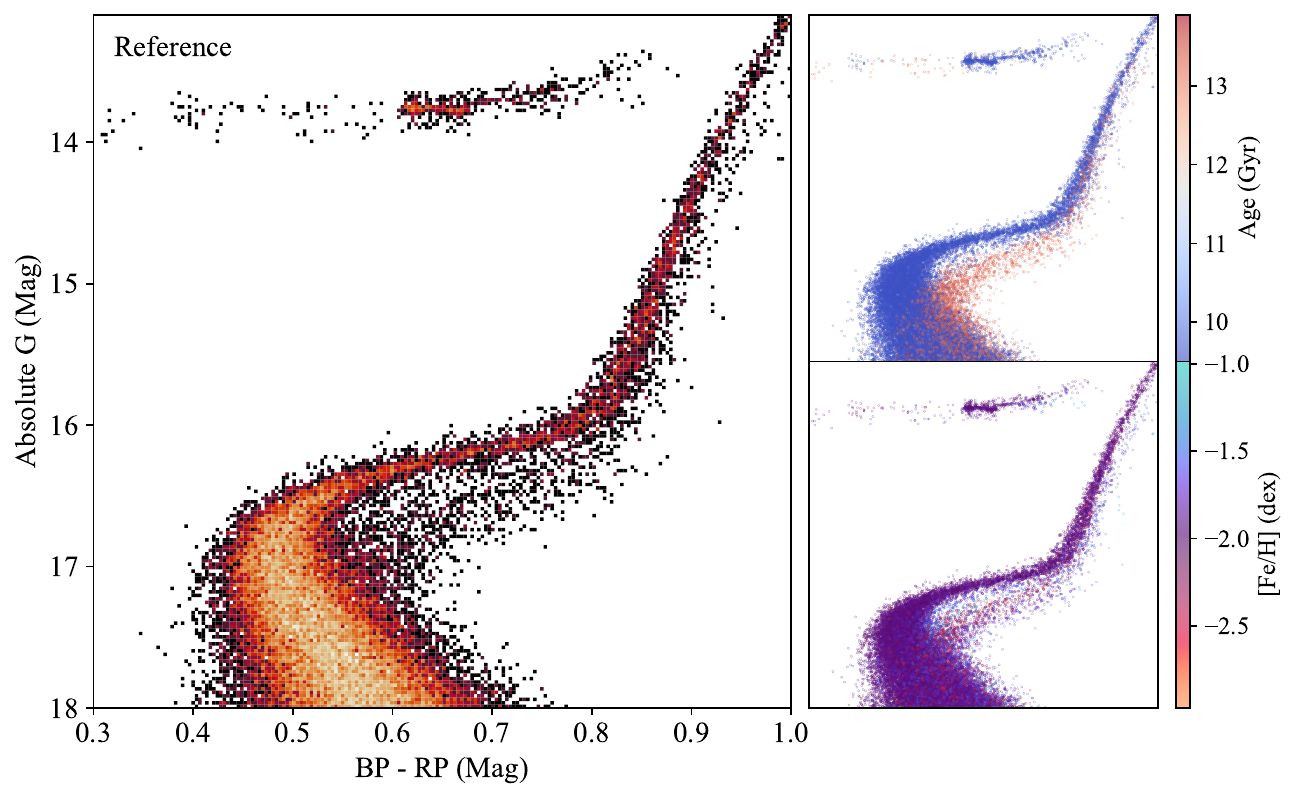}
    \caption{Mock CMDs of the Reference simulation coloured by density (left), stellar age and [Fe/H] (right). The synthetic photometric data was generated using the \texttt{py-ananke} \citep{2023arXiv231202268T} code using star particle data within 400\,pc of the galaxy's centre. The Gaia photometric system is used with no added photometric errors. The two stellar populations within the EDGE dwarf result in two clear main-sequence turnoffs, primarily due to a distinct separation in age.}    \label{fig:ref-cmd}
\end{figure*}

Figure \ref{fig:sfh} shows how these different assembly histories impact the star formation histories for each of our simulations (top panel) and the gas mass within 1\,kpc (bottom panel). In all cases, there is a pre-reionisation star formation. However, the Reference, Different Seed, and Later runs are all then quenched by reionisation. Their star formation only reignites again following a major merger after reionisation (as marked by the crosses; for more information about these mergers, see Table \ref{tab_dwarf_nsc}). By contrast, the Earlier simulation (orange) reaches a high enough mass at reionisation that it can continue to form stars almost continuously through to the present day. For this reason, it does not have a significant second starburst. As a result, unlike the other simulations, it does not form a NSC (see section \ref{sec:nuc-emerge}).

The bottom panel shows how gas builds up due to the major merger in each simulation. For the Reference, Different Seed and Later simulations, this excites a significant starburst shortly after the merger that then expels most of the inner gas, quenching star formation for the remainder of the simulation. By contrast, in the Earlier run this does not occur. The inner gas mass remains approximately constant for a Hubble time and star formation proceeds steadily for the duration of the simulation.

\subsection{Observational Properties}

In this section, we characterise our NSCs in more detail, presenting mock color-magnitude diagrams (\S\ref{sec:simCMD}), metallicities and abundances (\S\ref{sec:simchem}), and stellar kinematics (\S\ref{sec:simkinematics}).

\subsubsection{Colour Magnitude Diagrams}\label{sec:simCMD}

In Figure \ref{fig:ref-cmd}, we show mock colour-magnitude diagrams (CMDs) for the Reference dwarf, coloured by both density (left), star age and [Fe/H](right). To generate these, we used the \texttt{py-ananke} \citep{2023arXiv231202268T} code which utilises star particle data from cosmological simulations to generate synthetic photometric data. We select data (including positions, velocities, ages, masses and [Fe/H]) for stars within radii $<400$\,pc of the galaxic centre. We use the Gaia photometric system and do not include any photometric uncertainties within our CMD plots.

The Reference NSC has two main sequence turnoffs that owe, primarily, to a distinct separation in age for the pre- (${\sim}$13\,Gyrs) and post- ($\sim$9.5\,Gyrs) reionisation stellar populations (bottom panel). While the pre-reionisation population contains more metal-rich stars, and the post-reionisation contains more metal-poor stars, these have a more minor impact on the CMD. The other NSCs behave similarly and so we omit these for brevity. A key result of our model is that all NSCs formed in this way should contain pre- and post-reionisation populations in their CMDs and, therefore, at least two distinct main sequence turnoffs.

As we move outwards from the dwarf's centre, the younger starburst population becomes less prominent and the main-sequence associated with the, more spatially extended, older pre-reionisation stellar population increasingly dominates the CMD. We discuss this further, and how some nearby GCs exhibit similar behaviour, in section \ref{sec:discussion}.

\begin{figure*}
	\includegraphics[width=\textwidth]{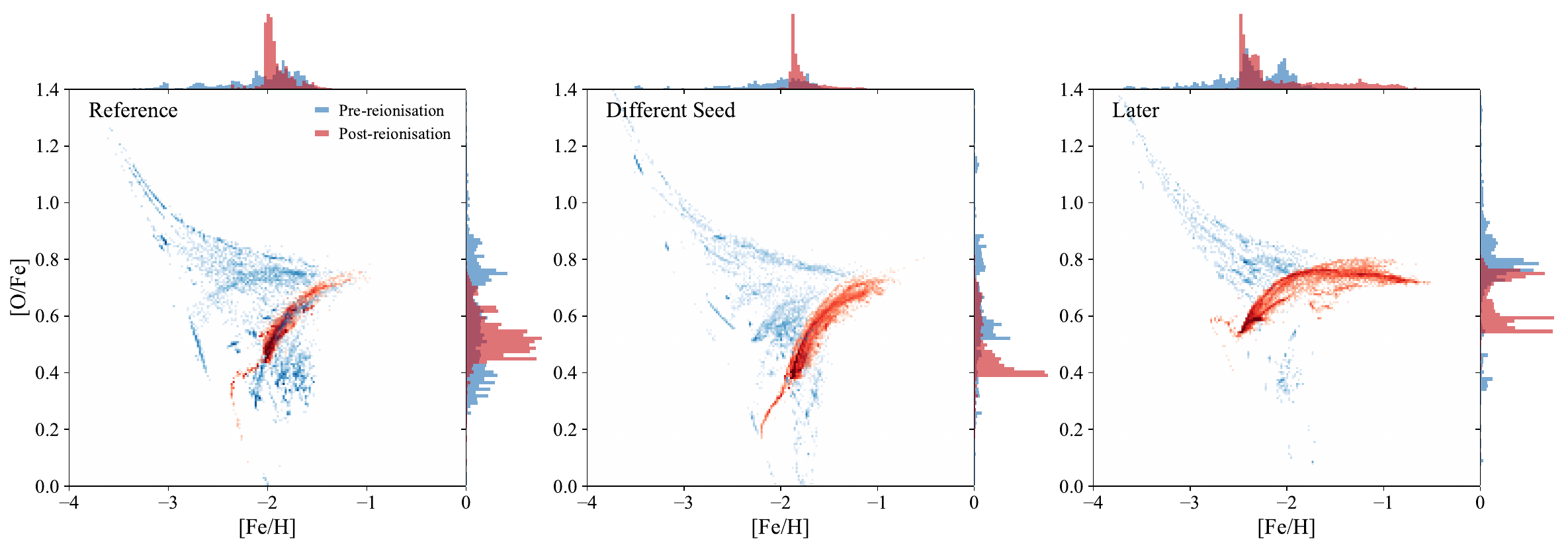}
    \caption{Abundance ratios of oxygen relative to iron as a function of [Fe/H] for the stars in all NSC-forming dwarfs. The pre- and post-reionisation populations are colored by blue and red respectively. [Fe/H] and [O/Fe] distributions for both populations are displayed above and to the right of the centre plot respectively. The pre-reionsation population displays an anticorrelation in contrast to the younger post-reionisation population.}
    \label{fig:chemical1}
\end{figure*}
 
\subsubsection{Chemical composition}\label{sec:simchem}

EDGE also allows us to characterise the oxygen and iron abundances of our simulated dwarfs and NSCs. In Figure \ref{fig:chemical1}, we show the abundance ratios of oxygen relative to iron as a function of [Fe/H], as well as the distributions of both [Fe/H] and [O/Fe], for each of our simulated nucleated dwarfs. The pre- and post-reionisation stellar populations are shown in blue and red respectively. Overall, there is an anticorrelation between [O/Fe] and [Fe/H] due to Type Ia supernovea, during and after the pre-reionization star formation, enriching the to-be-star-forming gas with iron and lowering the [O/Fe]. This is consistent with metal-complex clusters where iron-rich stars tend to be depleted in oxygen \citep[e.g.][]{bastian2018multiple}. However, our EDGE dwarfs display a positive correlation for the post-reionisation population at more metal-rich [Fe/H] values. This is likely due to the starburst occurring too quickly for Type Ia supernova to significantly contribute, thus causing the gas to rapidly reach high [O/Fe] found in typical core-collapse supernova ejecta.

All [Fe/H] distributions contain a significant peak, with a tail towards higher values that enhances the average iron abundance values. Such [Fe/H] distributions are similar to those observed in massive nearby GCs like Omega Centauri (OC; eg. \citealt{johnson2010chemOC}). While the peaks of the Reference and Different Seed are at similar metallicities, around -2.0\,dex, the Later simulation peaks at an [Fe/H] value closer to -2.5\,dex. This is expected because the Later simulation forms fewer stars before reionsation (see Figure \ref{fig:sfh} bottom panel) and, therefore, has a lower [Fe/H] before its post-reionization starburst. The [Fe/H] values are systematically lower than expected from literature trends for local volume dwarfs of similar magnitudes \citep{simon2019faintest}. This is most likely due to a combination of lower supernova yields \citep{woosley1995evolution} and the lack of an explicit model for radiative transfer which produces less violent outflows, increasing metal retention \citep[e.g.][]{agertz2020edge}. The Different Seed simulation has the highest iron peak value and the smallest spread. This owes to its extremely short, violent, starburst as compared to the other simulations -- see Figure \ref{fig:sfh}. As all of our simulated nucleated dwarfs have only one [Fe/H] peak, the split populations that naturally arise in our new NSC formation mechanism are not discernible from [Fe/H] measurements alone. 

Our dwarfs display complex [O/Fe] distributions, with some containing multiple peaks -- even within just the post-reionisation population. This is due to the sensitivity of oxygen abundances to the local star formation conditions. All simulations display at least one epoch of rapid enrichment during the starburst event, causing a sudden increase of $\sim$0.2\,dex in [O/Fe]. The Reference and Different Seed simulations contain multiple such epochs of rapid enrichment, each distinct in time and space. This leads to a broader [O/Fe] spread. The post-reionisation population for the Later simulation displays three clear peaks. The two peaks at $<0.6$\,dex correspond to spatially distinct star forming regions that have distinct [O/Fe]. Stars forming from the gas enriched by these two star forming regions then produces the third, broader, [O/Fe] peak at ${\sim}0.7$\,dex.

\begin{figure*}
	\includegraphics[width=0.95\textwidth]{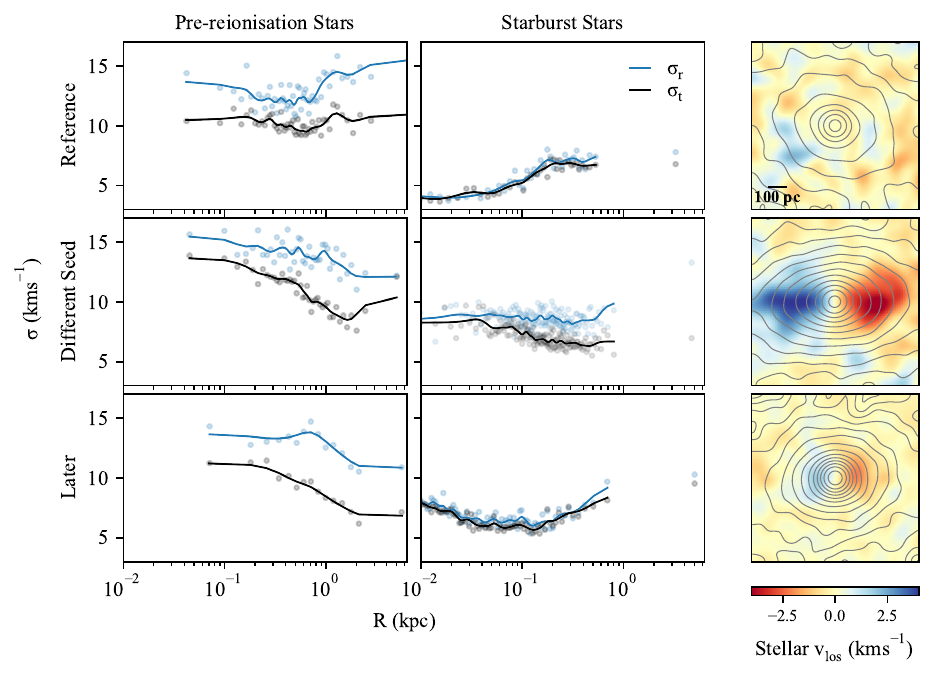}
    \caption{The radial (blue) and tangential (black) stellar velocity dispersions for the Reference, Different Seed and Later simulations (at z=0) for the stars formed before reionisation quenching (left; $<2.5$\,Gyrs from the start of the simulation), and stars formed during the starburst (right; $>2.5$\,Gyrs from the start of the simulation). In both cases, we align the dwarf `face on' such that the direction of the angular momentum vector of stars $<200$\,pc from the dwarf's centre is along the $z$-axis. Each radial bin contains an equal number of $\sim$ 200 star particles. The third column shows projected velocity maps of the Reference, Different Seed and Later nucleated dwarfs at $z=0$ covering an area of 1\,kpc. For each radial bin the average stellar velocity in the $z$-direction was found for all stars within the virial radius of the simulated dwarfs. Each dwarf was aligned `side on'. The contours represent stellar densities. The Different Seed simulation displays the strongest rotation, with a rotational velocity that reaches 4\,kms$^{-1}$ at 100-200\,pc from the dwarf's centre. The Later simulation shows a much weaker rotation, while the Reference simulation has no detectable rotation signal.}
    \label{fig:vel_disp}
\end{figure*}

\subsubsection{Stellar kinematics}\label{sec:simkinematics}
Figure \ref{fig:vel_disp} shows the radial and tangential velocity dispersions as a function of radius for the stars created before reionisation quenching (formed before 2.5\,Gyrs from the start of the simulation) and the stars created within the NSC-forming starburst (formed after 2.5\,Gyrs from the start of the simulation). Notice that the older stellar population is kinematically hotter and more radially anisotropic than the starburst NSC population.

The Different Seed simulation shows significant rotation. This is shown in Figure \ref{fig:vel_disp} (right panels) that plots the line-of-light velocities of all stars, oriented `side-on'. The Later simulation also shows evidence of rotation. In both cases, the rotation signal becomes most prominent (rising above an amplitude of ${\sim}2$\,km/s) at distances of $\sim$100-200\,pc from the dwarf's centre. The causes of rotation, and why some NSCs rotate strongly and others do not, will be explored further in future work. 

\section{Discussion}\label{sec:discussion}

\subsection{Model Limitations}

In this section, we discuss the dependency of our results on the sub-grid physics choices in the EDGE simulations. Our NSC formation mechanism relies on the complete quenching of star formation from cosmic reionisation. The mass-scale at which this occurs is sensitive to the detailed balance between heating and cooling after reionisation (e.g. \citealt{rey2020edge, benitez2020detailed} and references therein). Cooling will be enhanced in dwarf galaxies with higher metal retention, for example. This can occur due to the inclusion of explicit photo-ionization feedback that enhances metal retention (\citealt{agertz2020edge}), or due to reasonable changes in the stellar yield model \citep[e.g.][]{pillepich18}. Furthermore, key heating terms are uncertain, including the strength of the UV background after reionisation (UVB; e.g. \citealt{khaire2019new, puchwein2019consistent, faucher2020cosmic}) or the importance of stellar feedback from evolved binary stars (\citealt{rey2020edge}). Nonetheless, we argue that all these uncertainties only act to shift the absolute mass-scale at which our NSC formation mechanism occurs. For example, a stronger UVB than is assumed here would push reionisation-driven quenching, and therefore the mass scale at which dwarfs can form NSCs via this mechanism, to higher mass. (Indeed, merger-driven nuclear starbursts are well established in galaxy formation theory \citep{springel2005merge, renaud2022merger}, highlighting their robustness to sub-grid physics choices.)

Despite our EDGE simulations being able to resolve individual supernovae explosions, the simulations presented in this work do not account for radiative feedback from young stars. Since photo-ionisation and photo-heating from stars are active as soon as they are born, this feedback channel helps regulate starbursts and makes feedback less bursty and explosive in dwarfs \citep{agertz2020edge, smith2021efficient}. It remains to be seen how the nuclear starburst forming the NSC would proceed when including these additional physical effects. We will quantify this in future work. 

\begin{figure}
    \hspace{-4mm}\includegraphics[width=0.5\textwidth]{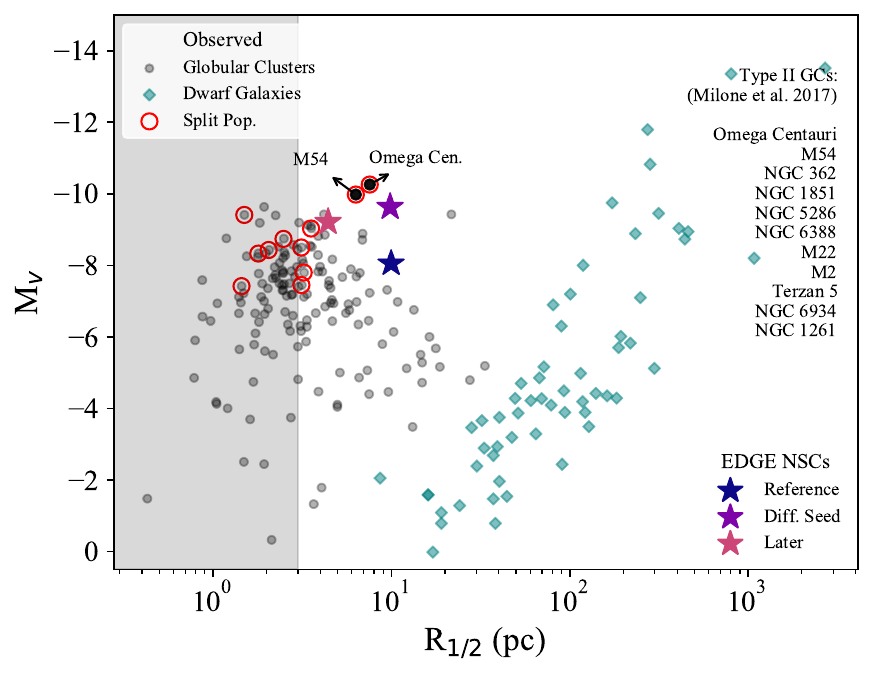}
    \caption{V-band magnitude as a function of the half-light radius. The EDGE NSC data is displayed in large star symbols - the V-band magnitude for these has been calculated from the stars within 4 $\times$ NSC half-light radii. Observational data for both globular clusters \citep{harris2010new, cerny2023delve, crnojevic2016deep, laevens2015sagittarius} and dwarf galaxies \citep{simon2019faintest, sand2022tucana, collins2022pegasus, mcquinn2023pegasus, conn2018nature, collins2023pisces} are shown in grey circles and teal diamonds, respectively. Globular clusters identified as Type II from \citealt{milone2017hubble} are circled in red and listed on the right of the plot. These include the clusters OC and M54 \citep{siegel2007acs}, which are directly labelled, in addition to Terzan 5 \citep{ferraro2009cluster}, NGC 1851, M22, NGC 5286, M2, NGC 362, NGC 1261, NGC 6388 and NGC 6934 \citep{piotto2012hubble, milone2017hubble}. The grey band to the left of the plot marks the simulation spatial resolution of 3pc.}
    \label{fig:Mv_rhalf}
\end{figure}

\subsection{A new mechanism for forming NSCs}

We have presented a new mechanism for forming NSCs (with stellar mass $\rm{M}_* \sim 0.3-1.3 \times 10^6$\,M$_\odot$) in low mass dwarf galaxies ($\rm{M}_{\rm{r}200\rm{c}} \sim 5 \times 10^9$\,M$_\odot$). Our model makes a number of key observational predictions:

\begin{itemize}
    \item The NSCs should have (at least) two distinct stellar populations with a large separation (${\sim 1}$\,Gyr) in age. This yields a CMD with two (or more) distinct main sequence turnoffs.
    \item The younger NSC population should be more metal rich than the older population.
    \item The younger NSC population should be kinematically colder than the older population.
    \item The younger NSC population should be more spatially compact than the older population.
\end{itemize}

\noindent
Ideally, we should compare our EDGE model predictions to isolated nucleated dwarfs in low density environments. Indeed, our model predicts that massive NSCs can form in dwarfs with stellar masses in the range $\sim 3-10 \times 10^6$\,M$_\odot$. We are beginning to explore such nucleated dwarfs in low density environments through new surveys like MATLAS \citep{MATLAS2021}. However, the best-studied nucleated dwarfs to date are in cluster environments, for example, the Virgo cluster (e.g. \citealt{NGVCsurvey2019}) and the Fornax cluster (e.g. \citealt{turner2012acs}). This creates additional complexity when making model comparisons because of the impact of tides, environment quenching and environmentally driven morphological transformations.

Several dwarfs in the Virgo cluster have a similar stellar mass to our dwarfs while also containing an NSC, with the smallest of these having a total stellar mass of M$_{\star}=10^{5.2} \rm{M}_{\odot}$ containing an M$_{\star}=10^{4.8} \rm{M}_{\odot}$ NSC \citep{NGVCsurvey2019}. The ratio of total galaxy stellar mass vs NSC stellar mass on average for these are higher than our dwarfs, indicating some tidal stripping of the host galaxy has likely decreased its total stellar mass. There are a couple which are more similar to our simulated dwarfs (e.g. NGVS J12:26:51.99+12:39:08.2 and NGVS J12:28:59.15+12:02:30.4), however, it is difficult to know for certain if these could be formed in a similar way to our dwarfs due to the denser environment. Similarly, both KKs58 (M$_{\star}=7.8 \times 10^{6} \rm{M}_{\odot}$) and KK197 (M$_{\star}= 4 \times 10^{7} \rm{M}_{\odot}$) within the Centaurus group have NSCs of stellar mass of M$_{\star} = 7.3 \times 10^{5} \rm{M}_{\odot}$ and $1.0 \times 10^{6} \rm{M}_{\odot}$, respectively \citep{fahrion2020metal}. However, in both cases the NSCs are more metal-poor ([Fe/H] = -1.75 $\pm$ 0.06 dex and -1.84 $\pm$ 0.05 dex for KKs58 and KK197, respectively) than the host galaxy ([Fe/H] = -1.35 $\pm$ 0.23 dex and -0.84 $\pm$ 0.12 dex for KKs58 and KK197, respectively). This could favour a GC inspiral scenario acting instead of, or in tandem, with the mechanism we present here, or point to continued star formation after the NSC forms that raises the metallicity of the host dwarf.

\subsection{Could some nearby GCs actually be accreted NSCs?}
Several studies in the literature have proposed that at least some GCs in the Milky Way might actually be NSCs accreted from now-dissolved nucleated dwarfs \citep[e.g.][]{freeman1993globular, boker2007globular}, with arguments ranging from some GC-NSC-candidates showing significant iron spreads \citep{pfeffer2021MW, dacosta2016nuclei}, to the presence of distinct populations in the colour magnitude diagrams \cite[e.g.][]{carretta2010m54+}. However, up to now it has not been explained why such complexity should necessarily imply that the object is a NSC.

Nearly all well-studied GCs exhibit chemical complexity that is referred to as `abundance anomalies' or `multiple stellar populations' \citep[e.g.][]{bastian2018multiple}. A smaller subset, however, show photometrically distinct main sequence turn-offs, as predicted by our new NSC model \citep{milone2017split}. Here, we suggest that these GCs are actually accreted NSCs.

One of these GCs, and also the most likely NSC candidate, is M54 which has long been thought to be the nucleus of the Sagittarius dwarf galaxy \citep[e.g.][]{carretta2010m54+, siegel2007acs,carlberg22}. M54 shows evidence for multiple split main-sequence turn-offs exactly as predicted by our model \citep[e.g.][]{alfarocuello19}. Similarly, OC has long been speculated to be a dwarf galaxy core disrupted from its host galaxy \citep{bekki2003formation, hilker2000omega, kuzma21}. It too shows at least three distinct main sequence turn-offs in its CMD \citep[e.g.][]{bellini10}.

Alongside these, there are many other `split main-sequence turnoff' GCs that also show a significant iron spread. We have undertaken a census and highlighted these in Figure \ref{fig:Mv_rhalf} which shows the V-band magnitude as a function of half-light radius for observed GCs (grey diamonds) and dwarf galaxies (green diamonds) as compared to our EDGE NSCs. Notice that our EDGE NSCs lie within the region occupied by bright GCs, particularly those identified as having clear split main sequence turn-offs. We speculate that many of these GCs could be accreted NSCs that formed through the new mechanism presented here. 

We can further test our NSC accretion hypothesis by comparing the spatial distributions, kinematics, and chemical compositions of the split populations in our model. Our model predicts that the pre-reionisation and starburst populations should be distinct, with the older stars occupying a separate, redder, main sequence (Figure \ref{fig:ref-cmd}). The older stars are also kinematically hotter (Figure \ref{fig:vel_disp}), more metal poor (Figure \ref{fig:chemical1}) and more extended than the younger NSC stars. It is challenging to compare these predictions in detail with nearby GCs like M54 and OC because nearby GCs: (i) have experienced significant tidal forces that we do not model in our EDGE simulations \citep[e.g.][]{kuzma21}; (ii) they undergo important two-body effects like mass segregation and evaporation that are not captured in EDGE \citep[e.g.][]{dehnen11,zocchi19}; and (iii) both M54 and OC have more than two main sequences and so have lived more complex lives than our simulated NSCs. Note that this latter point is not a problem for our model. Given the low halo mass in which our NSCs form, we can expect that they will accrete on to larger galaxies where they will sink to the centre, merging with other GCs and NSCs and possibly undergoing additional star formation through further gas cooling. Such processes will add further complexity to their observed properties today.

With the above caveats in mind, we now compare our NSC model to two well-studied GCs that are amongst the most likely to be NSCs: M54 and OC. M54 displays very similar qualitative behaviour to our model predictions. \citet{alfaro2020deep} report that the younger stars are kinematically colder and more metal rich than the older population, exactly as our model predicts, while \citet{alfarocuello19} show that the younger stars are more centrally concentrated than the intermediate-age stars (that they attribute to the host galaxy -- Sagittarius). The discussion surrounding OC is more complicated, with some studies agreeing that the apparently `older', redder, main sequence is more metal poor than the `younger', bluer, one \citep{piotto05,latour21}, while others appearing to show the opposite \citep{nitschai23, calamida17}. Some studies report more centrally concentrated metal-poor stars  \citep{calamida2020not} which contrasts with others who claim a central metal rich component \citep{vandeven06}. Many studies find that the multiple populations can be reasonably explained by a large spread in metallicity rather than age (e.g. \citealt{tailo2016mosaic-oc}).

Finally, we can look at rotation. Both OC and M54 show clear signs of rotation of magnitude 5-10\,km/s, particularly in the younger stellar populations \citep{merritt1996stellar,alfaro2020deep}. Some of our NSCs show a similar rotation amplitude (e.g. reaching $\sim$4 kms$^{-1}$ for the Different Seed simulation), but on larger radial scales ($100-200$\,pc as compared to $\sim$10\,pc for OC and M54 \citealt{merritt1996stellar,alfaro2020deep}). This difference may owe to the additional complexity seen in both M54 and OC as compared to our model. For M54 at least, there is a younger population that could have formed from further gas dissipation and collapse that could then be more highly rotating \citep[e.g.][]{alfaro2020deep}. However, it is important to note that we struggle to resolve rotation on sub-10\,pc scales given that our finest resolution cells are of size, 3\,pc. As such, we will return to this point with higher resolution simulations in future work.

\section{Conclusions}\label{sec:conclusions}

Using a suite of high-resolution simulations of dwarf galaxies drawn from the EDGE project ($\sim 3$\,pc spatial resolution), we have uncovered a new formation mechanism for NSCs. We find that massive NSCs ($M_*(<4\rm{r}_{1/2}) \sim 8 \pm 4 \times 10^5\,{\rm M}_{\odot}$) form in surprisingly low halo mass dwarf galaxies ($\rm{M}_{200} \sim 5 \times 10^9$\,M$_\odot$) at high redshift ($z \sim 2$). Since such low-mass dwarfs can accrete in large numbers onto larger galaxies, our mechanism could be the dominant mode of NSC seeding in galaxies of almost all mass. 

Our formation mechanism proceeds, as follows. First, the EDGE dwarf is sufficiently low mass that it is fully quenched by reionisation. It retains, however, a significant reservoir of ionised gas that is unable to form stars. Next, at some point after reionisation, the dwarf undergoes a major ${\sim}1:1$ merger. This compresses the substantial gas reservoirs in both the main dwarf and its merging companion, exciting rapid cooling and a major starburst. The NSC forms in the starburst that quenches star formation thereafter. This process naturally results in two stellar populations that have a $\sim$ billion year age-gap and mock colour-magnitude diagram (CMD) that resembles some nearby GCs like OC and M54 that have multiple, distinct, main sequence turnoffs.

We went on to study the detailed observational properties of our simulated NSCs, including their surface brightness, chemical composition, mock CMDs, and stellar kinematics. Our key results are as follows:

\begin{itemize}
  \item All of the surface brightness profiles for our nucleated EDGE dwarfs can be fit using two S\'{e}rsic profiles, one which corresponds to the NSC and the other to the host galaxy. 
  \item Mock CMDs of the inner stellar populations display clear split main sequence turnoffs due to the large ($\gtrsim 1$\,billion year) age separation in the pre- and post-reionisation stellar populations. Such split populations have already been observed in many bright GCs in the Milky Way. Our model suggests that these clusters, especially those which also have a significant [Fe/H] spread, could be accreted NSCs.
  \item The average [Fe/H] and [O/Fe] of our simulated NSCs are $\sim$ -2\, $\pm$ 0.1 dex and 0.5\, $\pm$ 0.1 dex, respectively. We qualitatively reproduce both the anti-correlation between iron and oxygen found in GCs that show significant [Fe/H] spread, and the overall shape of the [Fe/H] distribution.
  \item Our pre- and post-reionisation stellar populations are distinct, with the older pre-reionisation population being kinematically hotter, more radially anisotropic, slower rotating, and of lower chemical abundance than the younger population.

\end{itemize}

\section*{Acknowledgements}

The author would like to thank Dr Santi Cassisi and Dr Alex Riley for their support with synthetic CMD tools, as well as Madeleine McKenzie and Oliver Camilleri for the useful feedback and interesting discussions. 

This project has received funding from the European Union’s Horizon 2020 research and innovation programme under grant agreement No. 818085 GMGalaxies. JIR would like to thank the STFC for support from grants ST/Y002865/1 and ST/Y002857/1. ET acknowledges the UKRI Science and Technology Facilities Council (STFC) for support (grant ST/V50712X/1). MO acknowledges funding from the European Research Council (ERC) under the European Union’s Horizon 2020 research and innovation programme (grant agreement No. 852839). MR is supported by the Beecroft Fellowship funded by Adrian Beecroft.

This work also used the DiRAC@Durham (cosma6) facility managed by the Institute for Computational Cosmology on behalf of the STFC DiRAC HPC Facility (\url{www.dirac.ac.uk}). The equipment was funded by BEIS capital funding via STFC capital grants ST/P002293/1, ST/R002371/1 and ST/S002502/1, Durham University and STFC operations grant ST/R000832/1. DiRAC is part of the National e-Infrastructure.

\section*{Data Availability}
Data available upon reasonable request.



\bibliographystyle{mnras}
\bibliography{EDGE-NSCs} 






\bsp	
\label{lastpage}
\end{document}